\documentclass[prd,12pt,aps,superscriptaddress,double-spaced,floatfix,showkeys,showpacs]{revtex4}
\usepackage{graphicx,amsmath,bbm,bm,array,subfigure}
\usepackage[linktocpage=true,colorlinks=true,linkcolor=blue,citecolor=blue]{hyperref}

\begin{document}

\title {Hadron resonance gas with repulsive mean field interaction: Thermodynamics and transport properties}
\author{Guruprakash Kadam }
\email{guruprasadkadam18@gmail.com}
\affiliation{Department of Physics,
	Shivaji University, Kolhapur,
	Maharashtra-416004, India}
\author{Hiranmaya Mishra}
\email{hm@prl.res.in}
\affiliation{Theory Division, Physical Research Laboratory,
 Navarangpura, Ahmedabad - 380 009, India}

\date{\today} 

\def\be{\begin{equation}}
\def\ee{\end{equation}}
\def\bearr{\begin{eqnarray}}
\def\eearr{\end{eqnarray}}
\def\zbf#1{{\bf {#1}}}
\def\bfm#1{\mbox{\boldmath $#1$}}
\def\hf{\frac{1}{2}}
\def\sl{\hspace{-0.15cm}/}
\def\omit#1{_{\!\rlap{$\scriptscriptstyle \backslash$}
{\scriptscriptstyle #1}}}
\def\vec#1{\mathchoice
        {\mbox{\boldmath $#1$}}
        {\mbox{\boldmath $#1$}}
        {\mbox{\boldmath $\scriptstyle #1$}}
        {\mbox{\boldmath $\scriptscriptstyle #1$}}
}

\begin{abstract}
We discuss the interacting hadron resonance gas model to describe the thermodynamics of hadronic matter. 
While the attractive interaction between hadrons is taken care of by including all the resonances with zero width, 
the repulsive interactions between them are described by  a density-dependent mean field potential. 
The bulk thermodynamic quantities are confronted with the lattice quantum chromodynamics simulation results 
at zero as well as at finite baryon chemical potential. We further estimate the shear and bulk viscosity coefficients 
of hot and dense hadronic matter within the ambit of this interacting hadron resonance gas model.   
\end{abstract}

\maketitle

\section{Introduction}

Understanding the phase diagram of strongly interacting matter is one of the important and challenging  topics of 
current research in strong interaction physics- both theoretically and experimentally. The theoretical framework describing a nuclear matter
at a fundamental level  is quantum chromodynamics (QCD). At low temperature ($T$) and low baryon chemical potential ($\mu$) the
 fundamental degrees of freedom of QCD are colorless hadrons while at high temperature and high baryon density the fundamental degrees 
of freedom are colored quarks and gluons. Lattice quantum chromodynamics (LQCD) simulations at zero chemical potential
and finite temperature suggest a crossover transition 
 for QCD matter from a hadronic phase to a quark-gluon-plasma (QGP) phase\cite{Aoki:2006br,Borsanyi:2010bp,
Borsanyi:2015waa,Petreczky:2012fsa,Ding:2015ona,Friman:2011zz}. 
At zero chemical potential, the chiral crossover temperature is estimated to be $T_c\sim 156 $ MeV\cite{Bazavov:2018mes}.
While LQCD simulations at vanishing chemical potential has been quite successful, LQCD simulations at finite $\mu$ have been rather challenging particularly at high $\mu$ leading to large uncertainties in estimating the transition line in the $T-\mu$ plane of QCD phase diagram
\cite{Borsanyi:2012cr}. At small $\mu$, however, precise  computation of the transition line has been carried out recently
\cite{Bazavov:2017dus,Datta:2016ukp,Datta:2014zqa}.

The low energy effective models of QCD, viz.,the Nambu-Jona-Lasinio model\cite{Klevansky:1992qe,Hatsuda:1994pi},
 the quarak-meson coupling model\cite{Schaefer:2007pw} etc., provide a reasonable theoretical framework to explore
strongly interacting matter below the QCD transition temperature, $T_c$. These models are based on certain symmetries of QCD
and they are tremendously successful in describing many features of the QCD phase diagram at zero as well at finite baryon density. 
Apart from these symmetry-based models another model that has also been tremendously successful in describing the low temperature 
hadronic phase of QCD is the hadron resonance gas model (HRG). The hadron resonance gas model is the statistical model of 
QCD describing the low temperature hadronic phase of quantum chromodynamics. This model is based on the so-called 
 Dashen-Ma-Bernstein theorem which allows us to compute the partition function of the interacting system of hadrons in 
terms of scattering matrix\cite{Dashen:1969ep}. Using this S-matrix formulation of statistical mechanics it can be shown
 that if the dynamics of thermodynamic system of hadrons is dominated by narrow-resonance formation then the resulting 
system essentially  behaves like a noninteracting system of hadrons and resonances\cite{Dashen:1974jw,Welke:1990za,Venugopalan:1992hy}. 
This ideal HRG model, despite its success in describing hadron multiplicities in heavy-ion collisions
\cite{BraunMunzinger:1995bp,Yen:1998pa,Becattini:2000jw,Cleymans:1992zc,BraunMunzinger:2001ip,
Rafelski:2002ga,Andronic:2005yp,Chatterjee:2013yga,Chatterjee:2015fua}, fails to account for the short-range repulsive interactions 
between hadrons. It has been shown that the repulsive interactions modeled via excluded volume can have significant effect on 
thermodynamic observables, especially higher order fluctuations\cite{Albright:2015uua,Vovchenko:2016rkn,Vovchenko:2017zpj,Alba:2016fku} 
as well as  in the context of statistical hadronization\cite{BraunMunzinger:1999qy}.  One possible way to include these repulsive
 interactions is through van der Waals excluded volume procedure\cite{Rischke:1992rk,Singh:1991np}. Another approach is to treat 
the repulsive interactions in mean field way\cite{Kapusta:1982qd,Olive:1980dy}. Recently, the relativistic mean field approach 
has been used to calculate the fluctuations of conserved charges \cite{Huovinen:2017ogf}. 
This work discussed the repulsive mean field interactions which are present only at finite baryon density. They showed 
the deviations of higher order fluctuations estimated using the ideal HRG can be accounted by means of repulsive interactions 
treated in mean field way. The failure of ideal HRG model to explain the thermodynamical observable can be attributed 
to the fact that at high temperature and density the relativistic virial expansion up to the second-order virial coefficient 
cannot be a reasonable approximation and the validity of the HRG model needs to be checked against its agreement with LQCD results.

In the past few decades relativistic and ultrarelativistic heavy-ion collision experiments have  provided a unique
 opportunity to study the phase diagram of QCD. The relativistic hydrodynamics has been tremendously successful in 
simulating the evolution of matter created in HIC experiments~\cite{Gale:2013da,Schenke:2011qd,Shen:2012vn,Kolb:2003dz,
Teaney:2000cw,DelZanna:2013eua,Karpenko:2013wva, Holopainen:2011hq,Jaiswal:2015mxa}. In the relativistic hydrodynamic 
simulations the coefficients of shear and bulk viscosities influence various observables, viz.,the flow coefficients, 
the transverse momentum distribution of produced particles. In fact, it has been found that a finite but very small
 shear viscosity-to-entropy ratio ($\eta/s$) should be included in the hydrodynamic description to explain elliptic 
flow data\cite{Gyulassy:2004zy,Csernai:2006zz}. Further, $\eta/s$ obtained using AdS/CFT correspondence \cite{Kovtun:2004de} 
has put the lower bound on its value equal to $\frac{1}{4\pi}$ called the Kovtun-Son-Starinets (KSS) bound. This interesting 
finding has motivated many theoretical investigations to understand and rigorously derive this ratio
from a microscopic theory~\cite{Gavin:1985ph,Hosoya:1983xm,Prakash:1993bt,Dobado:2003wr,Itakura:2007mx,Chen:2007xe,
Dobado:2009ek,Demir:2008tr,Denicol:2013nua, Puglisi:2014pda,Thakur:2017hfc,Kadam:2018jaj,Kadam:2018hdo,Gorenstein:2007mw,
NoronhaHostler:2012ug}. The bulk viscosity coefficient ($\zeta$) has also been realized to be important to be 
included the dissipative hydrodynamics. During the expansion of the fireball, when the temperature approaches the 
critical temperature, $\zeta$ can be large and give rise to different interesting phenomena like cavitation when 
the pressure vanishes and the hydrodynamic description breaks down \cite{Rajagopal:2009yw, Bhatt:2011kr}. The effect of 
bulk viscosity on the particle spectra and flow coefficients have been investigated \cite{Monnai:2009ad,Denicol:2009am,Dusling:2011fd} 
while the interplay of shear and bulk viscosity coefficients have been studied in Refs. \cite{Song:2009rh,Noronha-Hostler:2013gga,
Noronha-Hostler:2014dqa}. The coefficient of bulk viscosity has been estimated for both the hadronic and the 
partonic systems~\cite{Dobado:2001jf,Davesne:1995ms,Kharzeev:2007wb,FernandezFraile:2009mi,Chen:2007kx,
NoronhaHostler:2008ju,Sasaki:2008um,FernandezFraile:2008vu,Dobado:2012zf,Ozvenchuk:2012kh,Gangopadhyaya:2016jrj,
Chakraborty:2010fr,Sasaki:2008fg,Berrehrah:2014ysa,Marty:2013ita,Samanta:2017ohm,Saha:2017xjq,Singha:2017jmq,Fraile2,SS_AHEP,
Deb:2016myz,Abhishek:2017pkp,Kadam:2015xsa,Khvorostukhin:2010aj,Dash:2019zwq,Zhang:2019uct,Mykhaylova:2019wci,Islam:2019tlo,
Ghosh:2018xll,Ghosh:2018nqi,Gao:2017gvf,Ghosh:2016clt,Attems:2016tby,Attems:2017zam,Florkowski:2017jnz}.

 Hydrodynamic simulation of the matter created in HIC collision requires information regarding equation of state (EoS) as well as 
the transport coefficients. In this work we analyze the QCD equation of state of hadronic matter at finite baryon chemical potential.
 We employ the hadron resonance gas model to estimate all the thermodynamic quantities. While the attractive interactions
 between hadrons are accounted for by including all the resonance states up to $2.25$ GeV, the short range repulsive interaction
 among hadrons are treated in the mean field approach. We call this model relativistic mean field hadron resonance 
gas model (RMFHRG). The RMFHRG differs from the Walecka type mean field models in the sense that in former the repulsive mean 
field interactions are present even at zero baryon density unlike the later case. We will also estimate the shear and bulk 
viscosity coefficient of hadronic matter within the ambit of RMFHRG. 

We organize the paper as follows.  In Sec.~\ref{secII} we compute the pressure and other bulk thermodynamic quantities for 
the interacting hadron resonance gas with a repulsive mean field interaction.   In Sec~\ref{secIII}, we discuss the results 
for the thermodynamics and confront them with the lattice simulation  results both at zero and finite chemical potential. We 
then estimate the viscosity coefficients for hot and dense hadronic matter within the ambit of the HRG model with mean field
 interactions. Finally, in Sec~\ref{secIV}, we summarize  the findings of the present investigation and give a possible outlook.

\section{Hadron resonance gas model with a repulsive mean field potential}
\label{secII}

Thermodynamic properties of hadron resonance gas model can be deduced from the grand canonical partition function defined as
\be
\mathcal{Z}(V,T,\mu)=\int dm [\rho_{b}(m)\: \text{ln}Z_{b}(m,V,T,\mu)+\rho_{f}(m)\: \text{ln}Z_{f}(m,V,T,\mu)]
\ee
where $\rho_{b}$ and $\rho_{f}$ are the mass spectrum of the bosons and fermions respectively.  We assume that the hadron mass spectrum is  given by
\be
\rho(m)=\sum_{a}^{\Lambda}g_{a}\delta(m-m_{a})\theta(\Lambda-m)
\ee
where $g_a$ is the degeneracy and $m_a$ is the mass of the $a$-{th} hadronic species. This discrete mass spectrum consists 
of all the experimentally known hadrons with cutoff $\Lambda$. One can set different cutoff values for baryons and mesons.

\noindent To include the effect of a repulsive interaction among hadrons, we use a repulsive mean field approach
as was used in Refs.  \cite{ Kapusta:1982qd,Olive:1980dy} 
and more recently in the case of baryons in Ref.\cite{Huovinen:2017ogf}.
In this approach, it is assumed that the repulsive interactions lead to a shift in the single particle energy and is given by
\be
\varepsilon_{a}=\sqrt{p^2+m_{a}^2}+U(n)=E_{a}+U(n)
\label{dispersion}
\ee
where $E_a=\sqrt{\zbf p^2+m_{a}^2}$ and $n$ is the total hadron number density. The potential energy $U$ represents repulsive 
interaction between hadrons, and it is taken to be a function of total hadron density $n$. For any arbitrary interhadron 
potential $V({\bf{r}})$, the potential energy is $U(n)=Kn$. Here, the phenomenological parameter $K$ is given by the integral 
of the potential $V(\bf r)$ over the spatial volume 
\cite{ Kapusta:1982qd,Olive:1980dy}.

%

In this work we assign different repulsive interaction parameter for baryons  and mesons. We denote the mean field parameter 
for baryons ($B$) and anti-baryons ($\bar{B}$) by $K_B$, while for mesons we denote it by  $K_M$. Thus,  for baryons (antibaryons)
\be
U(n_{B\{\bar{B}\}})=K_Bn_{B\{\bar{B}\}}
\label{potenbar}
\ee

and for mesons
\be
U(n_M)=K_Mn_M
\label{potenmes}
\ee

The total hadron number density is

\be
n(T,\mu)=\sum_{a}n_{a}=n_B+n_{\bar{B}}+n_M
\ee

where $n_{a}$ is the number density of $a$-th hadronic species. Note that $n_B$, $n_{\bar{B}}$  and $n_M$ are total baryon,
 antibaryon and meson number densities respectively. Explicitly, for baryons,

\be
n_{B}=\sum_{a\in B}\int d\Gamma_{a}\:\frac{1}{e^{(\frac{E_{a}-\mu_{\text{eff}}^a}{T})}+1}
\label{numdenbaryon}
\ee
where the sum is over all the baryons. Here,  $d\Gamma_{a}\equiv\frac{g_{a}d^{3}p}{(2\pi)^3}$, and, $\mu_{\text{eff}}^a=q^a\mu-U(n_{B})$
is a baryon effective chemical potential, with $q^a$ being the baryonic charge of $a$-th baryon and $\mu$ 
the baryon chemical potential. Similarly, for antibaryons

\be
n_{\bar{B}}=\sum_{a\in\bar{B}}\int d\Gamma_{a}\:\frac{1}{e^{(\frac{E_{a}-\bar{\mu}_{\text{eff}}^a}{T})}+1}
\label{numdenantbaryon}
\ee
 where $\bar{\mu}_{\text{eff}}^{a}=(\bar q^a\mu-U(n_{\bar{B}}))$ is an antibaryon effective chemical potential with $\bar q^a=-q^a$, 
which is the corresponding baryonic charge.   For mesons,

\be
n_{M}=\sum_{a\in M}\int d\Gamma_{a}\:\frac{1}{e^{\frac{(E_a-K_Mn_M)}{T}}-1}
\label{numdenmeson}
\ee
where the sum is over all the mesons. Note that $\mu=0$ for mesons since  the baryon charge is zero for them. In the Boltzmann 
approximation momentum integration can be readily performed and one can obtain much simpler expressions for the number
 density. For baryons we get

\be
n_{B}=\sum_{a\in B}\frac{g_{a}}{2\pi^2}m_{a}^2 T \mathcal{K}_{2}\bigg(\frac{m_{a}}{T}\bigg)e^{\frac{\mu^a_{\text{eff}}}{T}}
\label{numdenbar}
\ee

\be
n_{\bar{B}}=\sum_{a\in\bar{B}}\frac{g_{a}}{2\pi^2}m_{a}^2 T \mathcal{K}_{2}\bigg(\frac{m_{a}}{T}\bigg)e^{\frac{\bar{\mu}^a_{\text{eff}}}{T}}
\label{numdenantbar}
\ee
where $\mathcal{K}_{n}(z)$ is the modified Bessel function of order $n$. For mesons we get
\be
n_{M}=\sum_{a\in M}\frac{g_{a}}{2\pi^2}m_{a}^2 T \mathcal{K}_{2}\bigg(\frac{m_{a}}{T}\bigg)e^{-\frac{K_{M}n_{M}}{T}}
\label{numdenmes}
\ee

\noindent Equations (\ref{numdenbar} \ref{numdenantbar}\ref{numdenmes})  are self-consistent equations for number density which can be solved numerically.

The total baryon (antibaryon) energy density  is

\be
\epsilon_{B\{\bar{B}\}}=\sum_{a\in B\{\bar{B}\}}\int d\Gamma_{a}\:\frac{\varepsilon_{a}}{e^{\frac{[E_{a}-\mu^a_{\text{eff}}\{\bar{\mu}_{\text{eff}}\}]}{T}}+1}+\phi_{B\{\bar{B}\}}(n_{B\{\bar{B}\}})
\label{endenbaryon}
\ee

and for mesons

\be
\epsilon_{M}=\sum_{a\in{M}}\int d\Gamma_{a}\:\frac{\varepsilon_{a}}{e^{\frac{\varepsilon_{a}}{T}}-1}+\phi_{M}(n_M)
\label{endenmeson}
\ee
where $\phi(n)$ represents the correction to the energy density in order to avoid double counting the potential. 
It can be determined using the condition that $\varepsilon_{a}=\frac{\partial \epsilon}{\partial n_{a}}$. Taking the 
derivative of baryon energy density and using Eq. (\ref{potenbar}) we get

\be
\frac{\partial \phi_{B\{\bar{B}\}}}{\partial n_{B\{\bar{B}\}}}=-K_Bn_{B\{\bar{B}\}}
\ee

and hence

\be
\phi_B(n_{B\{\bar{B}\}})=-\frac{1}{2}K_Bn_{B\{\bar{B}\}}^2
\ee

Similarly for mesons one can obtain
\be
\phi_M(n_M)=-\frac{1}{2}K_Mn_M^2
\ee

Pressure of the  gas can now be readily obtained. For baryons

\be
P_{B\{\bar{B}\}}(T,\mu)=T\sum_{a\in B\{\bar{B}\}}\int d\Gamma_{a}\text{ln}
\bigg[1+ e^{-(\frac{E_a-\mu^a_{\text{eff}}\{\bar{\mu}^a_{\text{eff}}\}}{T})}\bigg]-\phi_{B\{\bar{B}\}}(n_{B\{\bar{B}\}})
\ee
and for mesons
\be
P_M(T)=T\sum_{a\in M}\int d\Gamma_{a}\text{ln}\bigg[1+ e^{-(\frac{\varepsilon_a}{T})}\bigg]-\phi_M(n_M)
\ee

Finally,  entropy density can be obtained from the fundamental thermodynamic relation $s=(\epsilon+P-\mu n)/T$. 
 It is worth noting that the effective interaction model we are considering is different from the relativistic Lagrangian model. In the
 latter case the repulsive mean fields are present only at nonzero baryon density, while in the 
former case the repulsive interactions are present even at zero baryon density.

\section{Results and discussion}
\label{secIII}

In the hadron resonance gas model it is customary to include all the hadrons and resonances up to  a certain cutoff $\Lambda$.
 We include all the mesons and baryons up to $\Lambda=2.25$ GeV  listed in the Ref.~\cite{Tanabashi:2018oca}. 
The phenomenological parameter $K$ is the spatially integrated value of the interhadron repulsive potential.
In Ref.\cite{Kapusta:1982qd} the potential was taken to be the same for all hadrons, i.e., for all the baryons and the mesons. 
In the present work we have taken this parameter different for mesons and baryons. For baryons we have taken it to be 
the same for all the baryons and the value is taken as in Ref. \cite{Huovinen:2017ogf}, 
 namely, $K_B$=450 MeV$fm^3$ for all baryons.
 Although different lattice calculations as well as chiral effective theories  indicate that, the strength
 can be different for nucleon-nucleon,
 hyperon-hyperon or nucleon-hyperon interactions, there is not enough information about hadrons to have a more realistic and 
sophisticated mean field model. For the mesons, we have taken a much smaller value
for the repulsion parameter $K_M$=50 MeV$fm^3$ in the present study.  The motivation in choosing these two phenomenological
parameters has been that the lattice results are reasonably reproduced regarding thermodynamics 
and then use them to estimate the viscosity parameters also at finite density.

\begin{figure}[t]
\vspace{-0.4cm}
\begin{center}
\begin{tabular}{c c}
 \includegraphics[width=8cm,height=8cm]{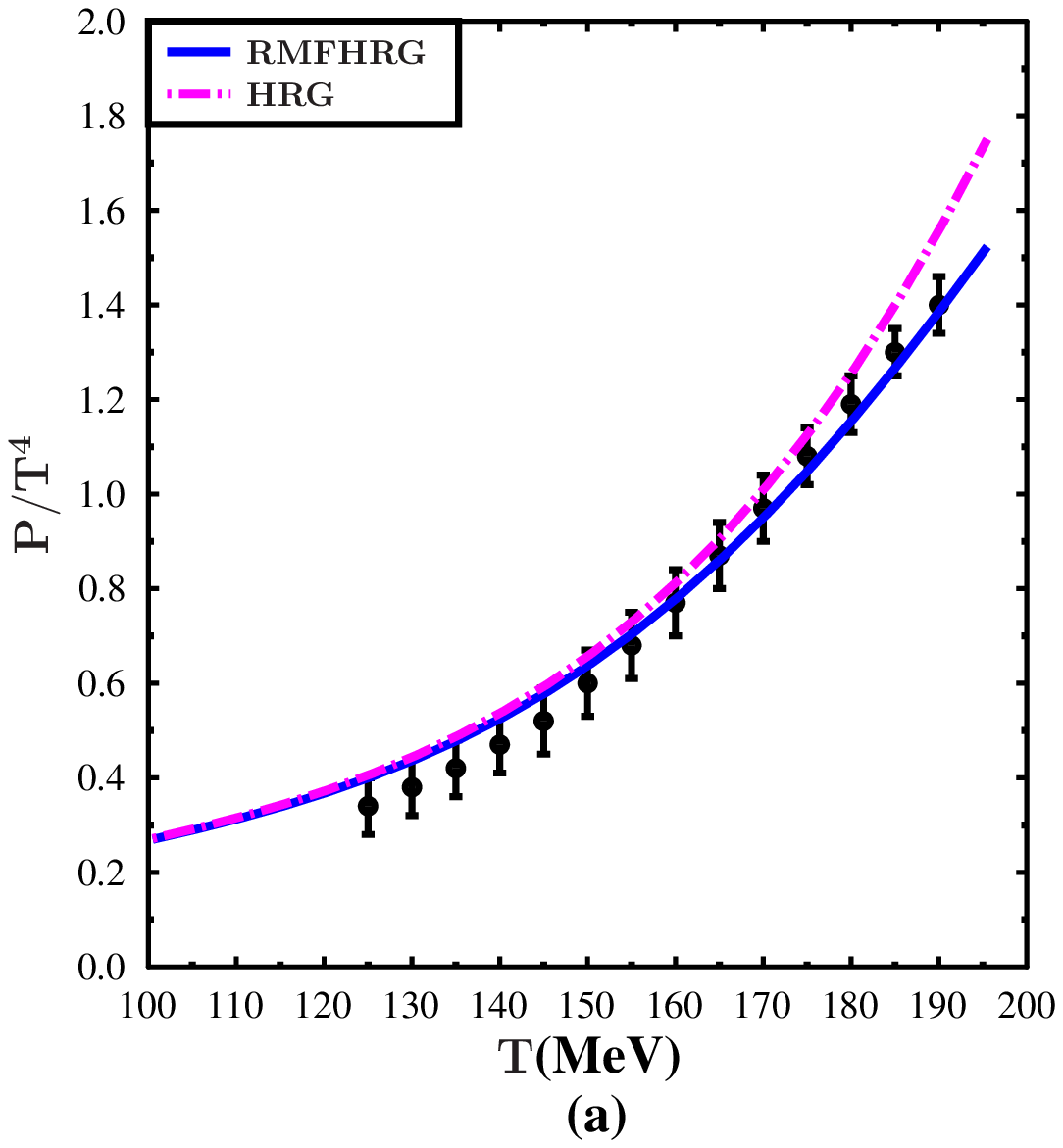}&
  \includegraphics[width=8cm,height=8cm]{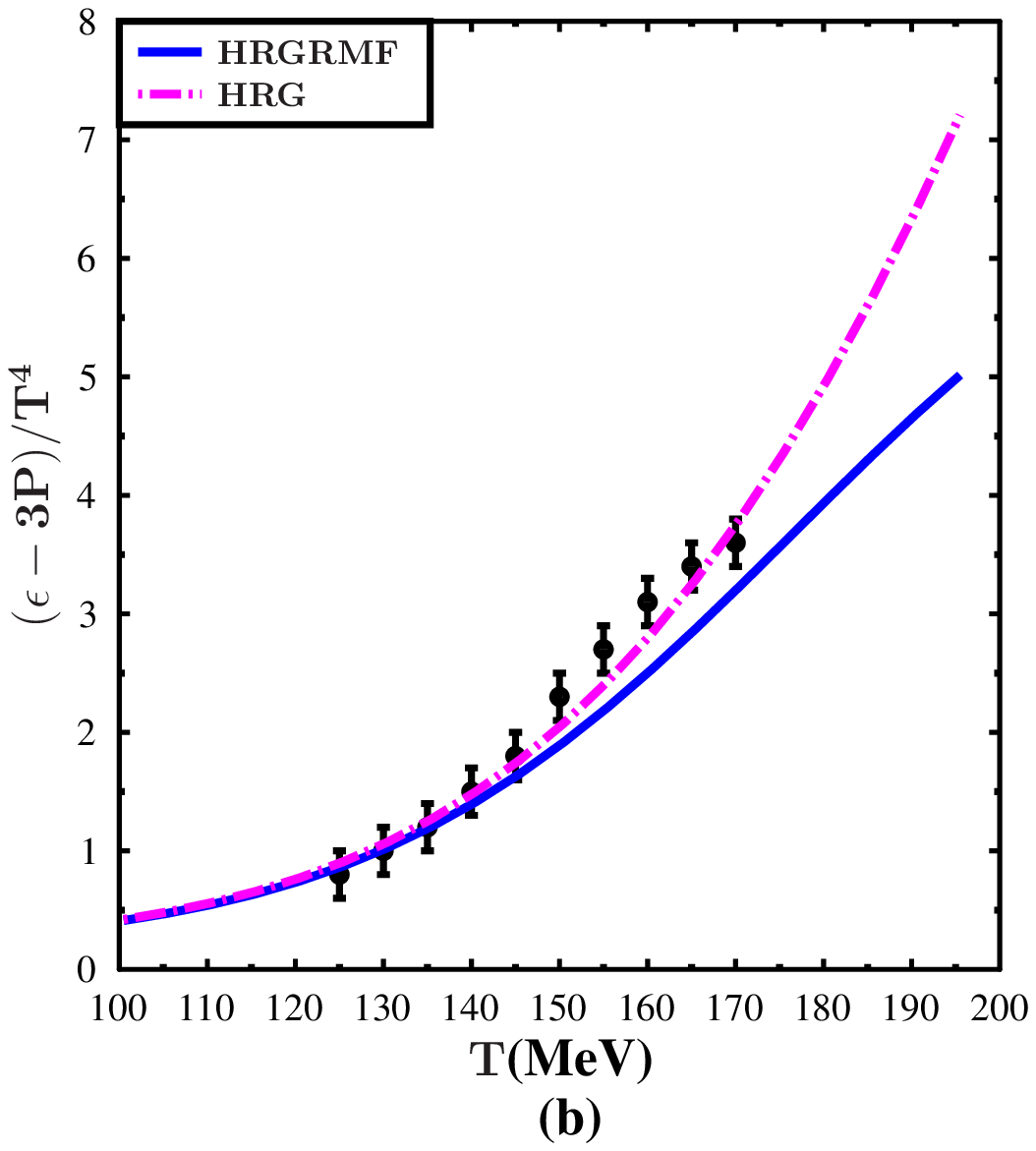}
  \end{tabular}
  \caption{ Scaled pressure (left panel) and the interaction measure (right panel) in RMFHRG and ideal HRG for $\mu=0$. The lattice data is taken from Ref.\cite{Borsanyi:2012cr}.} 
\label{EoS0}
  \end{center}
 \end{figure}

\begin{figure}[h]
\vspace{-0.4cm}
\begin{center}
\begin{tabular}{c c}
 \includegraphics[width=8cm,height=8cm]{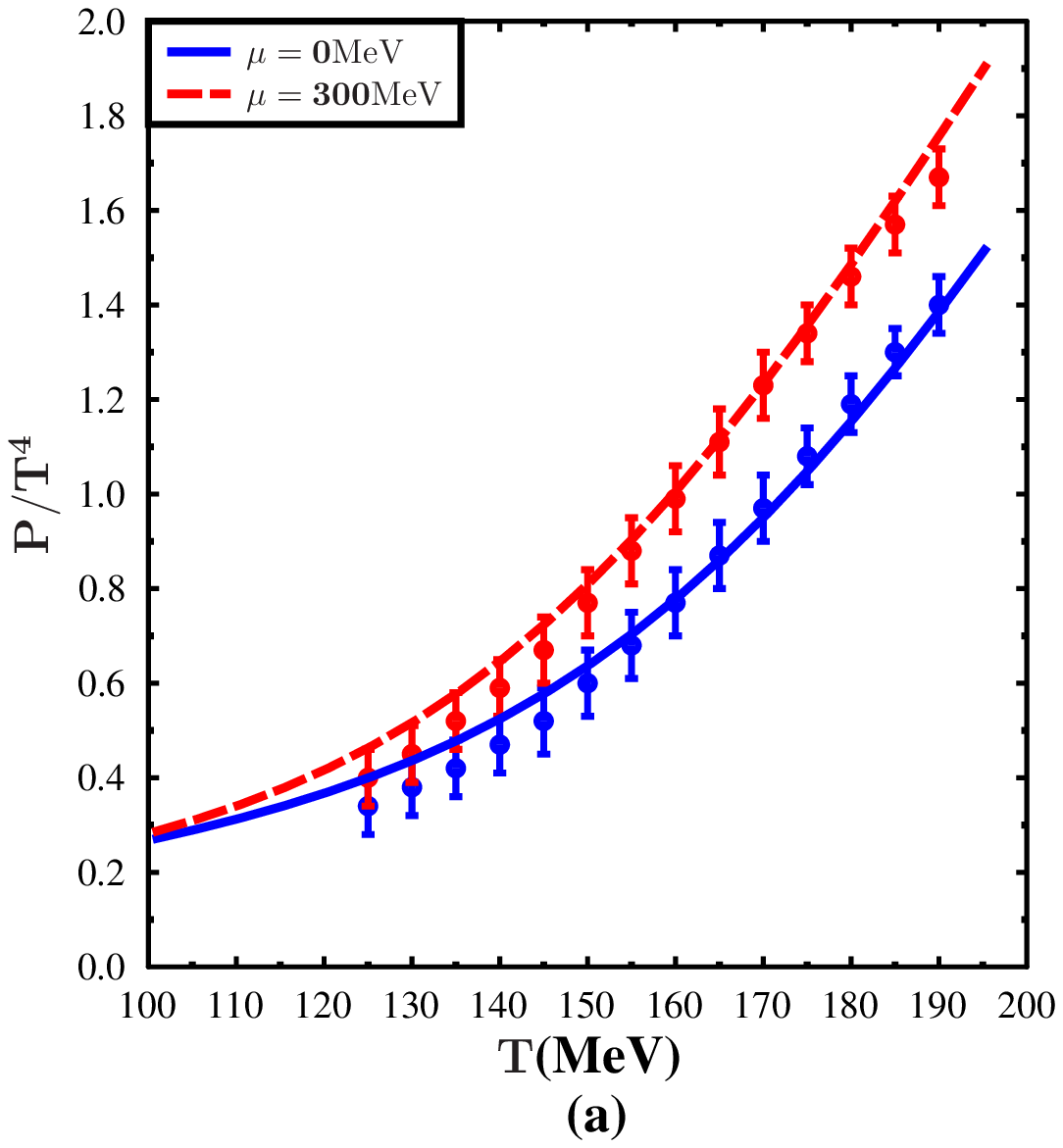}&
  \includegraphics[width=8cm,height=8cm]{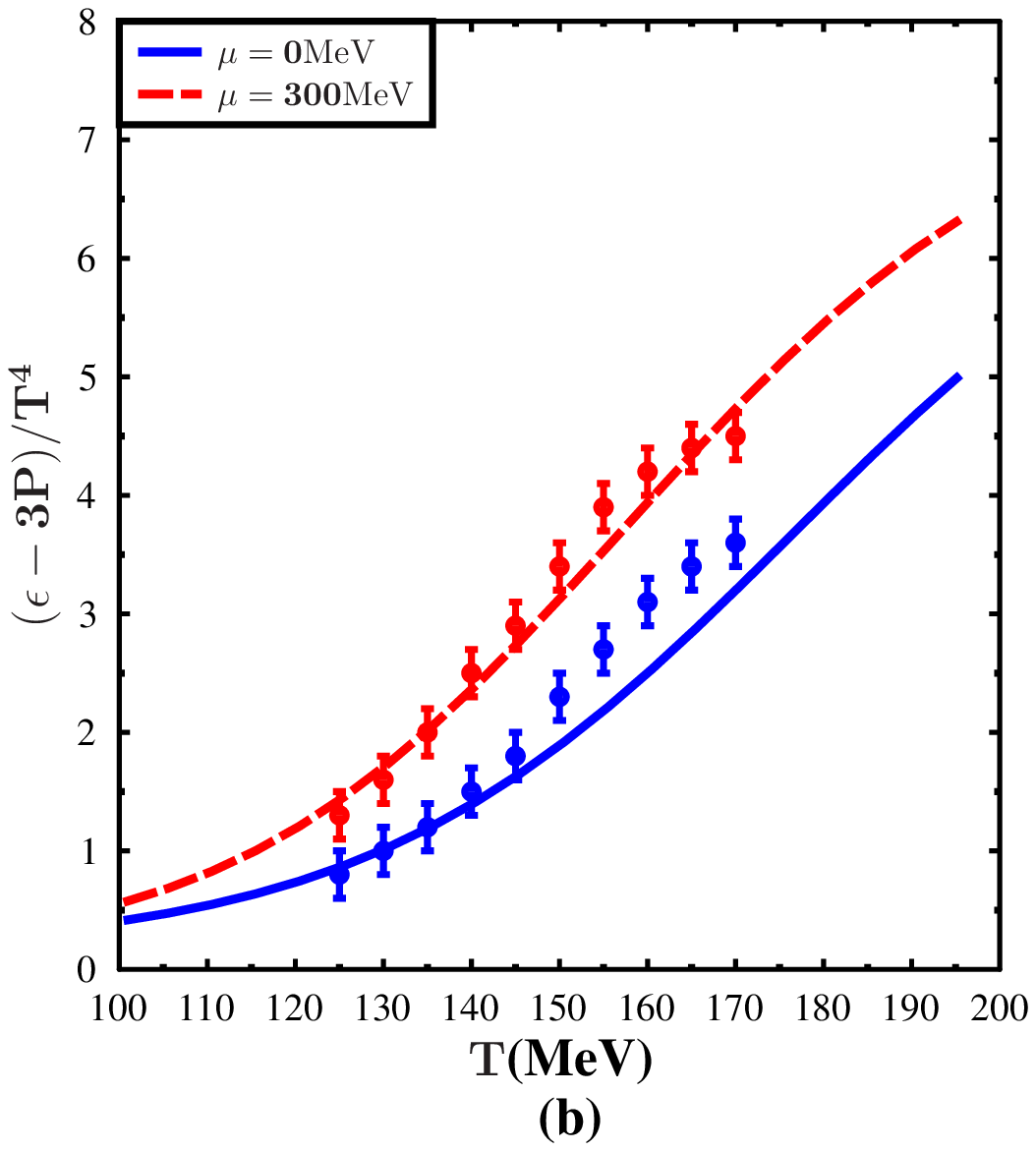}
  \end{tabular}
  \caption{ Scaled pressure (left panel) and the interaction measure (right panel) in the RMFHRG model at finite baryon chemical potential. The lattice data is taken from Ref.\cite{Borsanyi:2012cr}.} 
\label{EoS}
  \end{center}
 \end{figure}

  Figure \ref{EoS0} shows the scaled pressure and the interaction measure estimated within the  ambit of RMFHRG (blue solid curve) at
vanishing baryon chemical potential.
The dashed curve corresponds to ideal HRG results while the circles with error bars correspond to lattice QCD simulation 
results\cite{Borsanyi:2012cr}. We note that the effect of including the repulsive mean field interaction is to suppress the 
thermodynamical quantities compared to their ideal HRG estimation counterpart (dashed magenta curve). While the HRG pressure 
[Fig.\ref{EoS}(a)] starts to deviate from the lattice results at $T\sim 160$ MeV, the RMFHRG estimation agrees with the lattice 
results all the way up to 190 MeV. It is not reasonable to push the HRG model results above the QCD transition 
temperature ($T_c$) which LQCD estimates to lie in the range $155-160$ MeV\cite{Bazavov:2018mes}. The reason for this is twofold. 
First, the HRG approximation of the hadronic matter might break down at high density near and above $T_c$. Second, the hadrons 
do not exist above $T_c$. But a recent study\cite{Vovchenko:2016rkn} shows that the hadrons do not melt quickly as one
 would expect on the basis of ideal HRG model. In this study the authors have analyzed  the possible improvement of 
the ideal hadron resonance gas model by including the repulsive interactions between baryons. If one includes the attractive and 
the repulsive interactions between the baryons through van der Waals parameters, while keeping the meson gas ideal, the pressure 
of the hadron gas agrees with the LQCD data all the way above $T_c$. We may similarly conclude that the inclusion of repulsive 
mean fields might push the validity of the HRG model well above $T_c$. Nonetheless, we do not have any other strong reason to 
believe this  except for the apparent agreement with the LQCD results. 
  
  Unlike pressure the interaction measure is somewhat below LQCD results above $T=150$ MeV. It is an established fact that the 
socalled heavy Hagedorn states which are missing in our model contribute significantly to the energy density.
 The rapid rise of the energy density can be explained by extending ideal HRG  model by including continuum Hagedorn states alongwith
 the discrete states above the cutoff $\Lambda$ in the density of states\cite{NoronhaHostler:2008ju}.

\begin{figure}[h]
\vspace{-0.4cm}
\begin{center}
  \includegraphics[width=10cm,height=10cm]{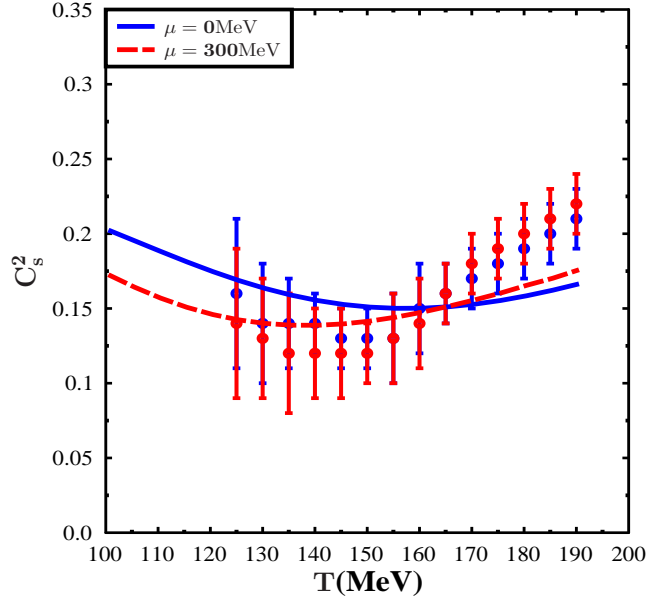}
  \caption{ Speed of sound in RMFHRG model at finite baryon density. The lattice data is taken from Ref.\cite{Borsanyi:2012cr}.} 
\label{cs2}
  \end{center}
 \end{figure}

 Figure \ref{EoS} shows the scaled pressure and interaction measure at finite baryon chemical potentials estimated within 
the ambit of RMFHRG. We note that the RMFHRG is in reasonable agreement with LQCD results even at finite baryon density. Further, 
the interaction measure is in better agreement with the LQCD results at finite density than in the $\mu=0$ case. However, while making 
this observation, we have to keep in mind that the lattice data of Ref.\cite{Borsanyi:2012cr}  is estimated at order $\mu^2$.
 Figure \ref{cs2} shows the adiabatic speed of sound at finite baryon density. The RMFHRG estimations of $C_s^2$ are
 within the errorbars of LQCD results. Further, the $C_s^2$ has a minimum at $T=155$ MeV for $\mu=0$ and at $T=140$ MeV 
for $\mu=300$ MeV which is in very close agreement with the LQCD results.

 The coefficients of shear and bulk viscosities can be extracted from the relativistic Boltzmann equation. These 
have been derived in Refs. \cite{Gavin:1985ph,Hosoya:1983xm} in the absence of any mean fields. While various 
authors have used different types of mean fields  to include medium effects as well as interactions, 
and have derived the transport coefficients, a rigorous, thermodynamically consistent derivation for the expressions
for different transport coefficients was derived in Ref.\cite{Albright:2015fpa},
both in the presence of a scalar and vector mean field.
The scalar mean field affect the mass while the repulsive vector mean field affects the chemical potential. 
The potential considered here does not affect the masses of the hadrons and is like a repulsive vector field, 
its effect is manifested in the effective chemical potential.
 In the relaxation time approximation of the Boltzmann equation, the shear ($\eta$) and bulk viscosity ($\zeta$) coefficients  are given by
\cite{Gavin:1985ph,Hosoya:1983xm,Albright:2015fpa}  
 
\be
\eta=\frac{1}{15T}\sum_{a}\int \frac{d^{3}p}{(2\pi)^{3}}\frac{{p}^{4}}{E_{a}^{2}}
({\tau}_{a}f_{a}^{0}+{\bar\tau}_{a}\bar f_{a}^{0})
\label{shearmulti}
\ee

\bearr
\zeta&=&\frac{1}{T}\sum_{a}\int \frac{d^{3}p}{(2\pi)^{3}}\bigg\{\tau_{a} f^{0}_{a}\bigg[E_{a}C_{n_{\mathcal{B}}}^{2}+\bigg(\frac{\partial P}{\partial n_{\mathcal{B}}}\bigg)_{\epsilon} -\frac{ p^{2}}{3E_{a}}\bigg]^{2}\nonumber\\&+&\bar\tau_{a} \bar f^{0}_{a}\bigg[E_{a}C_{n_{\mathcal{B}}}^{2}-\bigg(\frac{\partial P}{\partial n_{\mathcal{B}}}\bigg)_{\epsilon_a}-\frac{ p^{2}}{3E_{a}}\bigg]^{2}\bigg\}
\label{blkmulti}
\eearr
 
\noindent where $f^{0}$  is the equilibrium distribution function with an effective chemical potential
including the mean field and $C_{n_{\mathcal{B}}}^2$ is the speed of sound at constant  baryon number density. 
Further, in Eqs. (\ref{blkmulti}) and (\ref{shearmulti}), $\tau_a$ is the relaxation time for $a$-th hadronic particle 
species, while the barred quantities corresponds to that of antiparticles. In this work we use the thermally averaged relaxation 
time  which for a given species $a$ is given by
 
 \be
 \tau^{-1}_{a}=\sum_{b}n_{b}\langle\sigma_{ab}v_{ab}\rangle.
 \ee
 
 \noindent In the above, the sum is over all particles ($b$) other than the particle $a$ with which the scattering takes place; 
 $\sigma_{ab}$ is the total scattering cross section for  the process $a(p_a)+b(p_b)\rightarrow c(p_c)+d(p_d)$
and $v_{ab}$ is the 
relative velocity given by 
\be
v_{ab}=\frac{\sqrt{(p_a\cdot p_b)^2-m_a^2m_b^2}}{E_aE_b}
\ee

Further, $n_b$ is the number density for particle species $b$ given, with $g_b$ as the corresponding
degeneracy factor, as
\be
n_b=\frac{g_b}{(2\pi)^3}\int d\vec p f_b(\vec p)\simeq \frac{g_bT^3}{2\pi^2}(\beta m)^2\mathcal{K}_2(\beta m)
\exp(\beta\mu_{eff}^b)
\ee
where the last step is written down in the Boltzmann approximation and $\mu_{eff}^b=\mu-K_Bn_B$ 
for baryons, $\mu_{eff}^b=\bar\mu-K_Bn_{\bar B}$  for antibaryons, and $\mu_{eff}^b=-K_M n_M$ for mesons.

Finally, the thermal average cross section $\langle\sigma_{ab}v_{ab}\rangle$ is given as
\be
\langle\sigma_{ab}v_{ab}\rangle=\frac{\int d^3p_ad^3p_b\sigma_{ab}v_{ab}f_a(p_a)f_b(p_b)}{\int d^3p_a d^3p_bf_a(p_a)f_b(p_b)}.
\label{sigv}
\ee

The only unknown quantity in Eq.(\ref{sigv}) is the total cross section. We estimate it as follows.
In Born approximation, the scattering amplitude $f(\theta,\phi)$  for a particle with mass $m$ that encounters a 
scattering potential $V(r)$ is given by \cite{griffiths}
 \begin{equation}
 f(\theta,\phi)=-\frac{m}{2\pi}\int d^3r\:V(r)=-\frac{mK}{(2\pi)}
\label{fthetaphi}
 \end{equation}
 and thus the cross section is given by 
\be
\sigma=4\pi \left(\frac{mK}{2\pi}\right)^2
\label{sig}
\ee
Then the thermal averaged cross section can be written as\cite{Cannoni:2013bza,Gondolo:1990dk}
\be
\langle\sigma_{ab}v_{ab}\rangle=\frac{\sigma}{8m_a^2m_b^2\mathcal{K}_2(\beta m_a)\mathcal{K}_2(\beta m_b)}\int_{(m_a+m_b)^2}^\infty dS
\frac{[S-(m_a-m_b)^2]}{\sqrt{S}}[S-(m_a+m_b)^2]\mathcal{K}_1(\beta\sqrt{S})
\ee
where $\sqrt{S}$ is the centre-of-mass energy.
Clearly, we have suppressed the baryon/meson index in the expression for the cross section for
 the parameter $K$ in Eq.(\ref{sig}). It may be relevant here to mention that, while the cross section is independent of 
temperature and chemical potential, the thermal averaged cross section $\langle\sigma v\rangle$, in general, depends upon
 temperature and chemical potential.  However, in the Boltzmann approximation $\langle\sigma v\rangle$ is independent of $\mu$. 
After evaluating the thermal averaged relaxation time
for each species, we estimate the viscosity coefficients using Eqs. (\ref{shearmulti})  and (\ref{blkmulti}).

\begin{figure}[h]
\vspace{-0.4cm}
\begin{center}
\begin{tabular}{c c}
 \includegraphics[width=8cm,height=8cm]{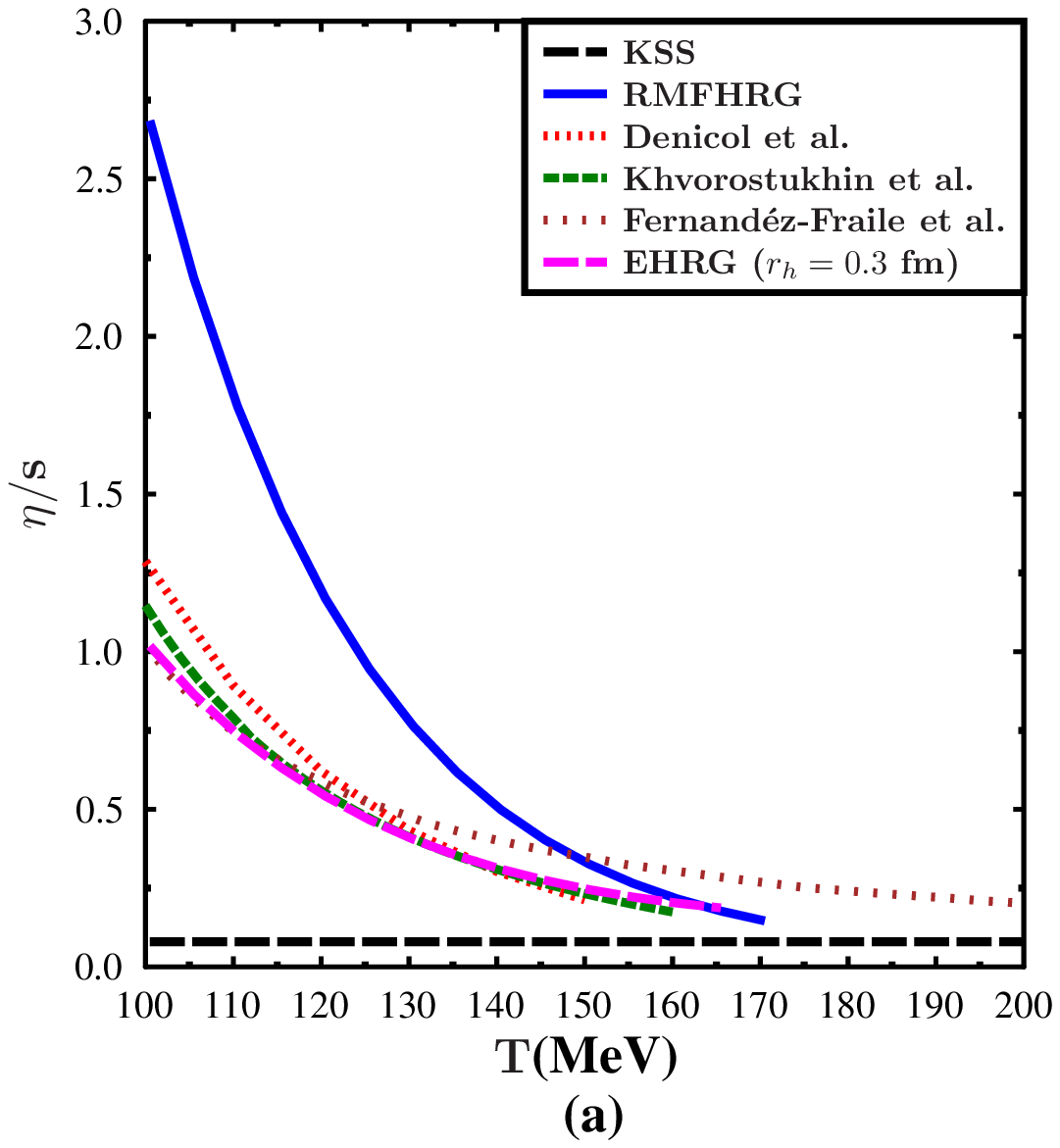}&
  \includegraphics[width=8cm,height=8cm]{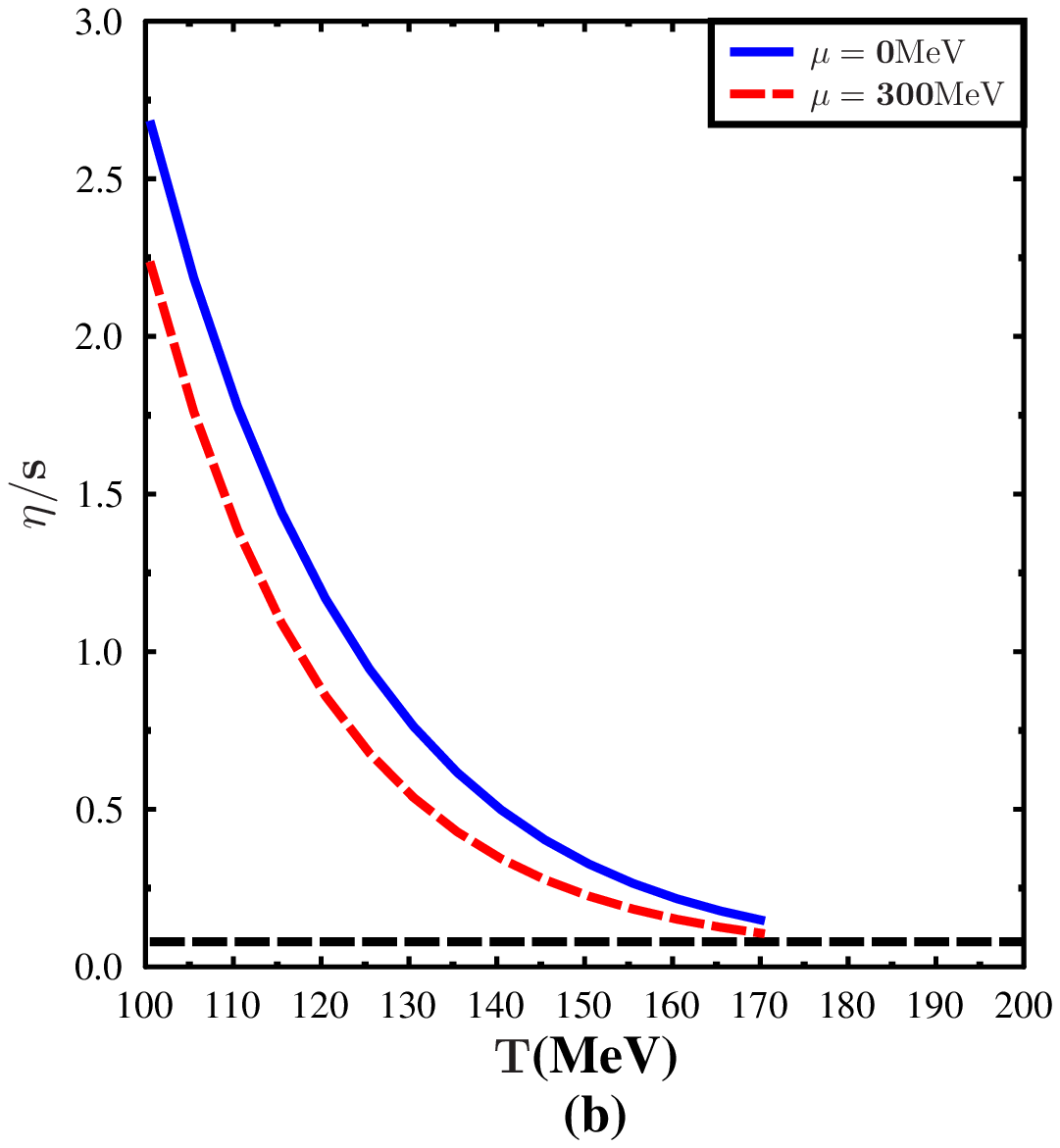}
  \end{tabular}
  \caption{ The left panel shows shear viscosity to entropy density ratio  estimated within RMFHRG and compared 
with other model estimations.
These figures correspond to $\mu=0$. The right panel shows $\eta/s$ for two different baryon chemical potentials.} 
\label{shearvisc}
  \end{center}
 \end{figure}
 
 Figure \ref{shearvisc} shows the ratio of shear viscosity to entropy density as a function of temperature. We have compared the 
ratio $\eta/s$ estimated within the ambit of RMFHRG with various other model calculations
\cite{Denicol:2013nua, Khvorostukhin:2010aj,Kadam:2015xsa,FernandezFraile:2009mi,Kovtun:2004de}. The red dashed curve corresponds 
to Chapman-Enscog method with constant cross sections\cite{Denicol:2013nua}. The dashed green curve corresponds to
the relativistic Boltzmann equation in the relaxation time approximation. The thermodynamic quantities in this model have
 been estimated  using the scaled hadron masses and coupling (SHMC) model\cite{Khvorostukhin:2010aj}. the brown dashed curve 
corresponds to estimations made using relativistic Boltzmann equation in RTA. The thermodynamic quantities are estimated within
the  excluded volume hadron resonance gas model (EHRG)\cite{Kadam:2015xsa}. The dot-dashed orchid curve corresponds 
to the $\eta/s$ of meson gas estimated using chiral perturbation theory\cite{FernandezFraile:2009mi}. While the ratio $\eta/s$ in our
 model is relatively large at low temperature as compared to other models it rapidly falls and approaches to the 
Kovtun-Son-Starinets (KSS) bound at $T\sim 170$ MeV.

 \begin{figure}[h]
\vspace{-0.4cm}
\begin{center}
\begin{tabular}{c c}
 \includegraphics[width=8cm,height=8cm]{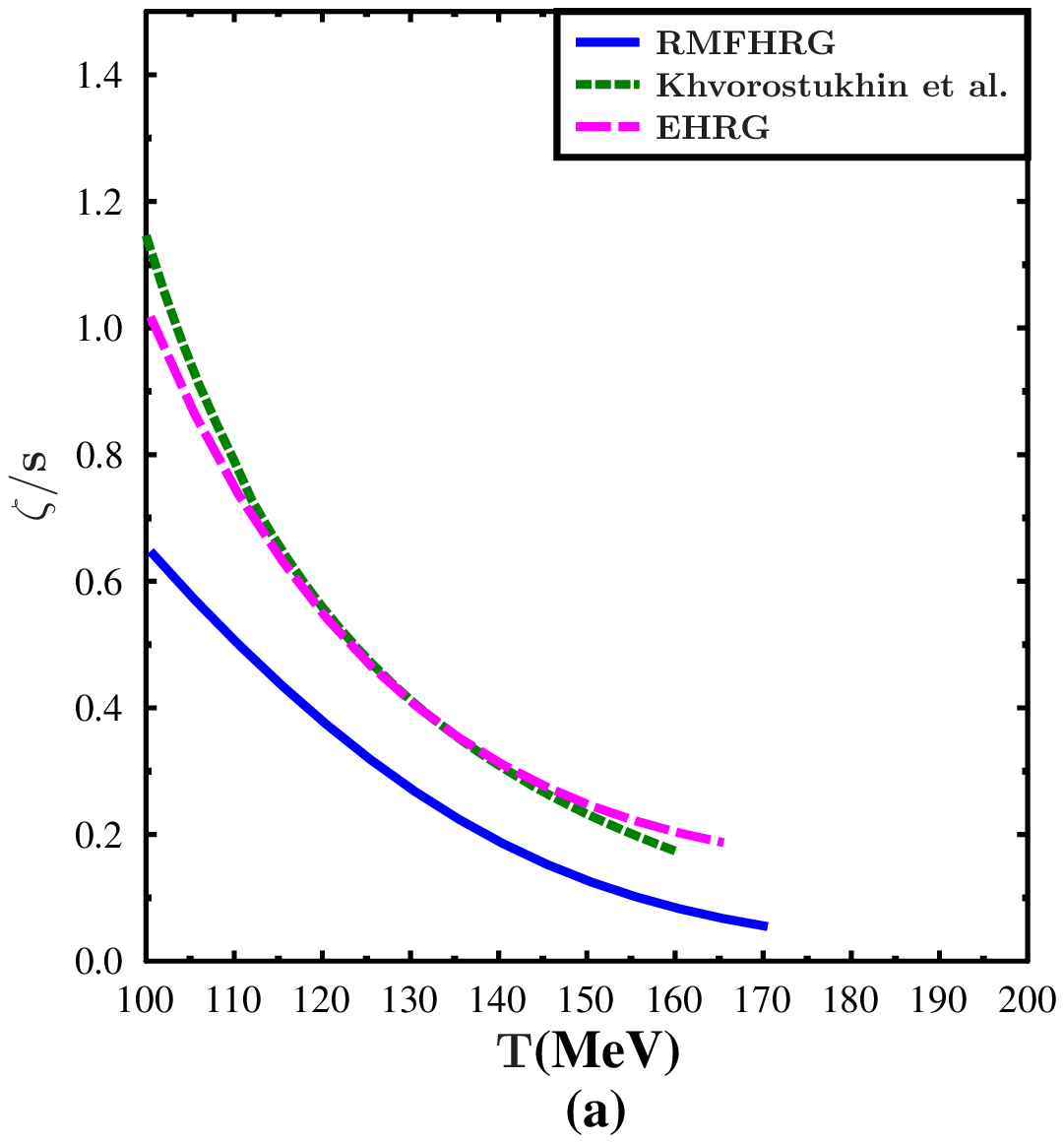}&
  \includegraphics[width=8cm,height=8cm]{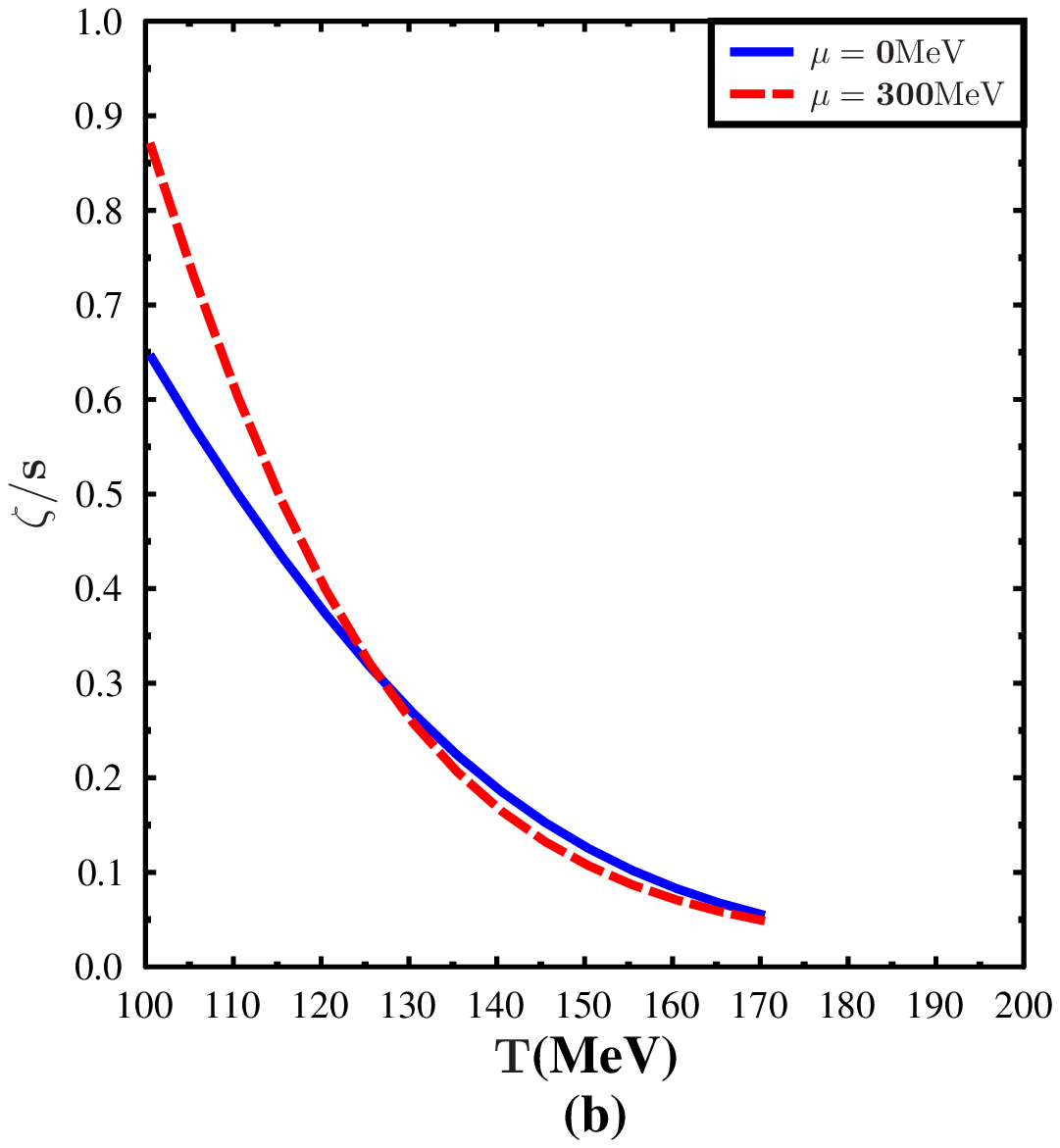}
  \end{tabular}
  \caption{The left panel shows bulk viscosity to entropy density ratio  estimated within RMFHRG and 
compared with other model estimations.
These results are for $\mu$=0. The right panel shows $\zeta/s$ at two different baryon chemical potentials.} 
\label{bulkvisc}
  \end{center}
 \end{figure}

Figure \ref{bulkvisc} shows the ratio of bulk viscosity to entropy density as a function of temperature.
 In Fig.\ref{bulkvisc}(a) the blue solid curve corresponds to the RMFHRG compared with that of the EHRG model (dashed magenta curve)
\cite{Kadam:2015xsa} and the SHMC model\cite{Khvorostukhin:2010aj}. Note that the ratio $\zeta/s$ is smaller when the 
repulsive interactions are treated in a mean field way. Figure\ref{bulkvisc}(a) shows the ratio $\zeta/s$ at finite 
baryon chemical potential. At low temperature the ratio is larger at finite $\mu$; it drops below that of $\mu=0$ case at high temperature.
 This observation may be attributed to the fact that the entropy density rises much faster than that of $\zeta$ itself at finite baryon density as compared to that of zero baryon density case.

\begin{figure}[h]
\vspace{-0.4cm}
\begin{center}
\begin{tabular}{c c}
 \includegraphics[width=8cm,height=8cm]{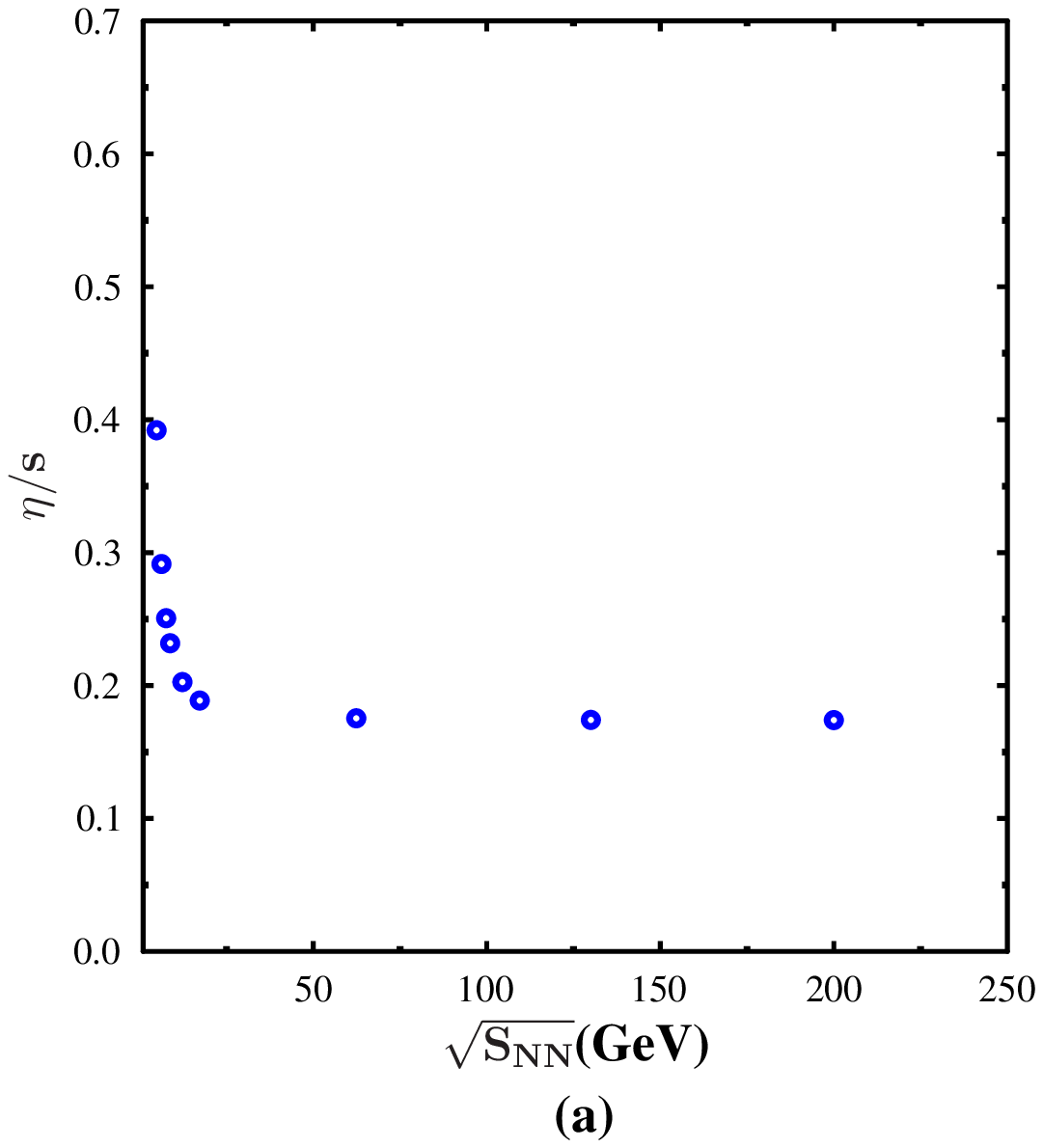}&
  \includegraphics[width=8cm,height=8cm]{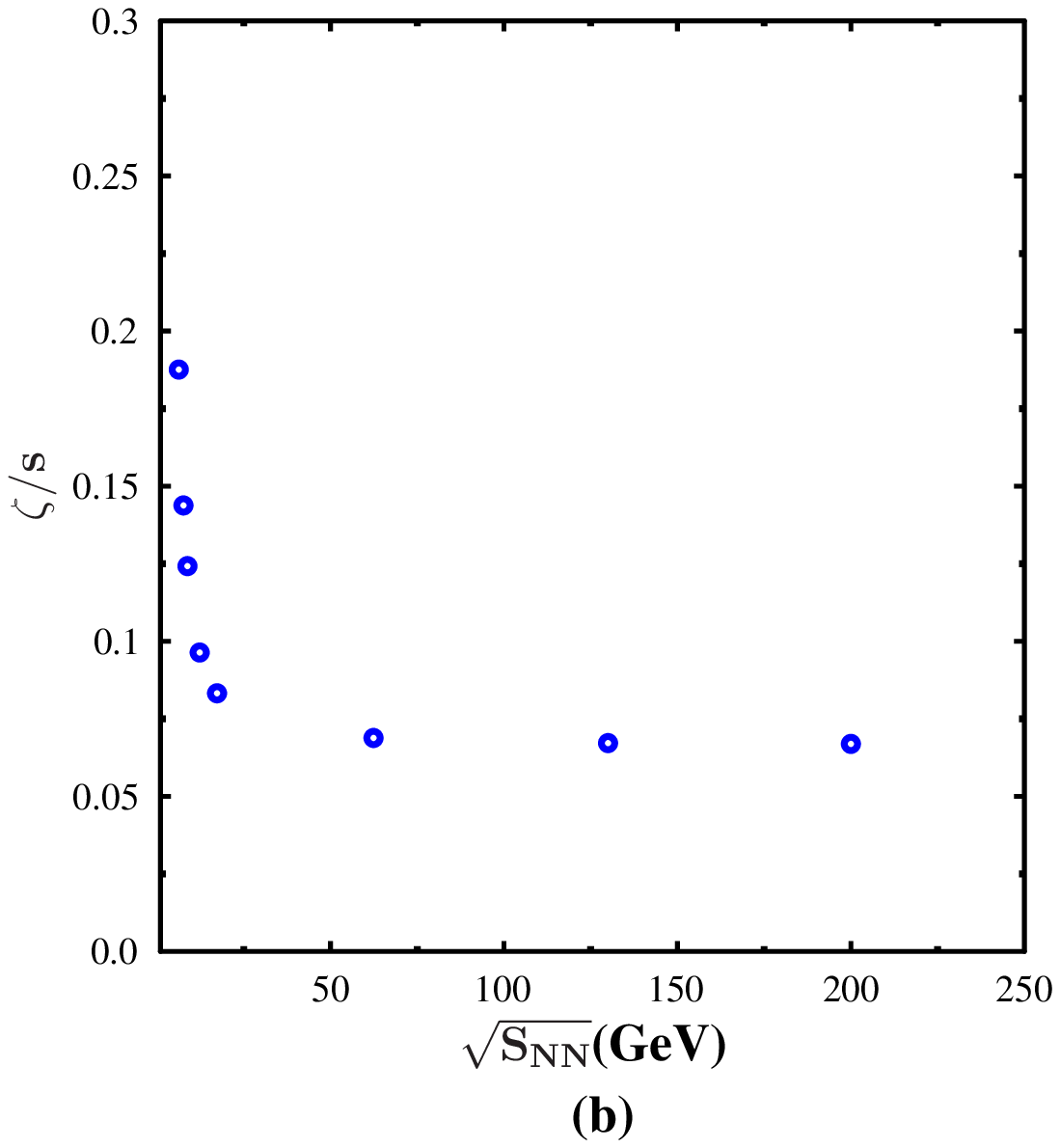}
  \end{tabular}
  \caption{ Viscosity coefficients along the freeze-out curve. The freeze-out parametrization is taken from Ref.\cite{Bugaev:2013jza}.} 
\label{snn}
  \end{center}
 \end{figure}

 In the context of heavy nucleon-nucleon (NN) collision experiments viscosity coefficients can be estimated along freezeout curve 
by finding the beam energy ($\sqrt{S_{NN}}$) dependence of the temperature and chemical potential. This is extracted 
from a statistical thermal model description of the particle yield at various $\sqrt{S_{NN}}$ 
\cite{Cleymans:2005xv,Bugaev:2013jza,Tawfik:2013dba}. We use the parametrization of the freezeout curve $T(\mu_{B})$  given in Ref.\cite{Bugaev:2013jza} as
 
 \be
 T(\sqrt{S_{NN}})=c_{+}(T_{10}+T_{20}\sqrt{S_{NN}})+c_{-}\bigg(T_{0}^{\text{lim}}+\frac{T_{30}}{\sqrt{S_{NN}}}\bigg)
 \label{freezT}
 \ee
 \be
 \mu(\sqrt{S_{NN}})=\frac{a_{0}}{1+b_{0}\sqrt{S_{NN}}}
 \label{freezemu}
 \ee
where, $T_{10} = -34.4$ MeV, $T_{20} = 30.9$ MeV/GeV, $T_{30} = -176.8$ GeV MeV, $T_{0}^{\text{lim}} = 161.5$ MeV, 
$a_{0}=1481.6$ MeV and $b_{0}=0.365$ GeV$^{-1}$.  The functions $c_{+}$ and $c_{-}$ smoothly connect the different 
behaviors of centre-of-mass energies. 
 
 Figure(\ref{snn}) shows viscosity coefficients, $\eta/s$ and $\zeta/s$ along the freeze-out curve. It can be noted that 
the fluidity measure rapidly falls at low $\sqrt{S}$  and then it remains almost constant at higher $\sqrt{S}$ values. 
This indicates that the matter produced in  heavy-ion collision experiments with wide range of collision energies can
 exhibit substantial elliptic flow.

\section{Summary}
\label{secIV}
In this paper we confronted the RMFHRG model with LQCD at zero as well as finite density. The repulsive interaction between the
hadrons is treated using the mean field approach. The thermodynamic quantities estimated within RMFHRG are found to be in 
reasonable agreement with LQCD at zero as well as finite chemical potential. The agreement of interaction measure $\epsilon-3P/T^4$ 
estimated within RMFHRG is rather poor above $T=145$ MeV. In fact the interaction measure rises very rapidly near $T_c\sim 156$MeV. 
This rapid rise of energy density can be explained by extending ideal HRG model by including continuum Hagedorn states alongwith 
the discrete states. We used this RMFHRG equation of state to estimate the shear and bulk viscosity coefficients of hadronic matter.
 We found reasonable agreement of both the viscosity coefficients with previous results. The shear viscosity to entropy density 
ratio $\eta/s$ estimated within the RMFHRG is large at low temperature as compared to other calculations. 
This behavior is due to the smaller cross section of mesons in our model. But $\eta/s$ estimated in our calculation 
rapidly drops at high temperature and approaches the  KSS bound at $T\sim 170$ MeV. We further found that $\eta/s$ at 
finite chemical is smaller in magnitude as compared to that of zero chemical potential but the overall behavior as a function
 of temperature do not change. We also found the reasonable agreement of the ratio $\zeta/s$ with previous results. Finally, we 
have estimated viscosity coefficients along the freeze-out line. We found that both the ratios, $\eta/s$ and $\zeta/s$, 
attain constant values at high $\sqrt{S}$ values. This indicates that the matter produced in heavy-ion collision experiments with a
 wide range of collision energies can exhibit substantial elliptic flow.

\section{Acknowledgement}
GK is financially supported by the DST-INSPIRE faculty award under Grant No. DST/INSPIRE/04/2017/002293.

\end{document}